\documentclass[a4paper]{article}

\usepackage{amsfonts, amsmath, amssymb, amsbsy, graphicx}

\usepackage[sort,numbers]{natbib}

\usepackage{url} 

\begin{document}

\begin{center}
\textbf{\LARGE A search for new variable stars using digitized Moscow collection plates}
\end{center}

\begin{center}
\textbf{Kirill~Sokolovsky$^{1,2}$, Sergei~Antipin$^{2,3}$, Daria~Kolesnikova$^3$,
Alexandr~Lebedev, Nikolai~Samus$^{3,2}$, Lyudmila~Sat$^2$, Alexandra~Zubareva$^{2,3}$}
\end{center}

\begin{center}
{\it
\noindent $^1$Astro Space Center of Lebedev Phys. Inst.,
Profsoyuznaya Str. 84/32, 117997 Moscow, Russia \\
$^2$Sternberg Astronomical Inst., Moscow State Uni.,
Universitetskii~pr. 13, 119992 Moscow, Russia \\
$^3$Institute of Astronomy RAS, 48, Pyatnitskaya~Str.,
119017 Moscow, Russia \\ }
\end{center}

\begin{abstract}
By digitizing astronomical photographic plates one may extract full
information stored on them, something that could not be practically achieved
with classical analogue methods. We are developing algorithms for variable
objects search using digitized photographic images and apply them to 30\,cm
($10^\circ \times 10^\circ$ field of view) plates obtained with the 40\,cm astrograph in
1940--90s and digitized with a flatbed scanner. Having more than 100 such
plates per field, we conduct a census of high-amplitude ($>0.3m$) variable stars changing their 
brightness in the range $13<m<17$ on timescales from hours to
years in selected sky regions. This effort led to discovery of $\sim 1000$ new
variable stars. 
We estimate that $1.2 \pm 0.1$\,\% of all stars show easily-detectable
light variations; $0.7 \pm 0.1$\,\% of the stars are eclipsing binaries 
($64 \pm 4$\,\% of them are EA type, $22 \pm 2$\,\% are EW type and 
$14 \pm 2$\,\% are EB type); $0.3 \pm 0.1$\,\% of the stars are red variable
giants and supergiants of M, SR and L types.
\end{abstract}

\section{Introduction}

Historical sky photographs present a record of positions and
brightness of astronomical objects. They are used to study
behaviour of objects as diverse as Solar system bodies
\cite{2011MNRAS.415..701R,2013arXiv1310.7502K}, binary stars
\cite{2011PZ.....31....1S,2012AJ....144...37Z}, and
active galactic nuclei \cite{2010AJ....139.2425N,2013A&A...559A..20H} 
on timescales not accessible with CCD imaging data.
A few authors used digitized photographic plates to identify previously unknown
variable objects \cite{2001A&A...373...38B,2004A&A...428..925V,2008A&A...477...67H,2012ApJ...751...99T}.

The Moscow collection contains about 60000 photographic plates (mostly
direct sky images) dating back to 1895. The most important part of the
collection, known as the ``A'' series, are 22300 plates taken in 
1948--1996 with the 40\,cm astrograph \cite{2010ASPC..435..135S}. 
These are blue-sensitive 30\,cm by 30\,cm plates covering $10^\circ \times
10^\circ$ field on the sky down to the limiting magnitude of $B\sim17.5$.
The typical exposure time is 45\,min.

The original aim of obtaining the ``A'' series plates was to study
variable stars. We decided to extend this work using modern image
analysis techniques. The first tests confirmed that it is possible to find
variable objects using small sections of plates digitized with a flatbed
scanner
\cite{2006PZP.....6...18S,2006PZP.....6...34M,2007PZP.....7....3K,2007PZP.....7...24K}
and we went ahead to process a series of full-sized $10^\circ \times 10^\circ$
plates \cite{2008AcA....58..279K,2010ARep...54.1000K}.
Below we describe the current state of the project.

For the original tests we used a pair of CREO/Kodak EverSmart Supreme~II
scanners operating at 2540\,dpi resolution ($1.\!\!^{\prime\prime}2$/pix).
While showing good photometric performance (typically $<0.1m$ accuracy of an
individual measurement), the scanners were suffering from problems common to
many flatbed scanners including poor out-of-the-box astrometric performance
caused by irregular motion of the scanner drive (Fig.~\ref{fig:saw}) and stitches between image
parts digitized during different passes of the scanning array over a
photographic plate. It takes about 40~minutes to digitize a half of the 30\,cm plate with the Supreme~II
scanner. The time it takes to clean a plate and manually place it into a 
scanner is small compared to the scanning time. The original Supreme~II
scanners were recently replaced by the new Epson Expression~11000XL which
provides a factor of two increase in scanning speed operating at 2400\,dpi
resolution ($1.\!\!^{\prime\prime}4$/pix). The Supreme~II and
Expression~11000XL scanners provide comparable results in terms of
photometric and astrometric accuracy.
Still, because the scanning process is so slow, we consider it to be more of a technology 
development tool and an opportunity to investigate a few individual fields
rather than a practical way to digitize all the Moscow plate collection in reasonable time.

\section{Plate digitization and data reduction}

The plates are digitized with 48\,bit color depth (16\,bit/color channel)
and saved into a TIFF format using a control software supplied with a
scanner. TIFF is the only format capable of preserving such color depth that
is supported by the control software of both scanner types. This format
has a built-in limitation that a file cannot be larger than $2^{32}$
bytes (4\,Gb) corresponding to a $\sim 9400\times9400$\,pix image at
48\,bit color depth. This means that a 30\,cm plate cannot be scanned into
a single file, so in practice a plate is scanned into two files with a small
overlap between the two images. This does not pose a problem for the
subsequent analysis.

\begin{figure}
\minipage{0.48\textwidth}
\includegraphics[width=1.0\textwidth]{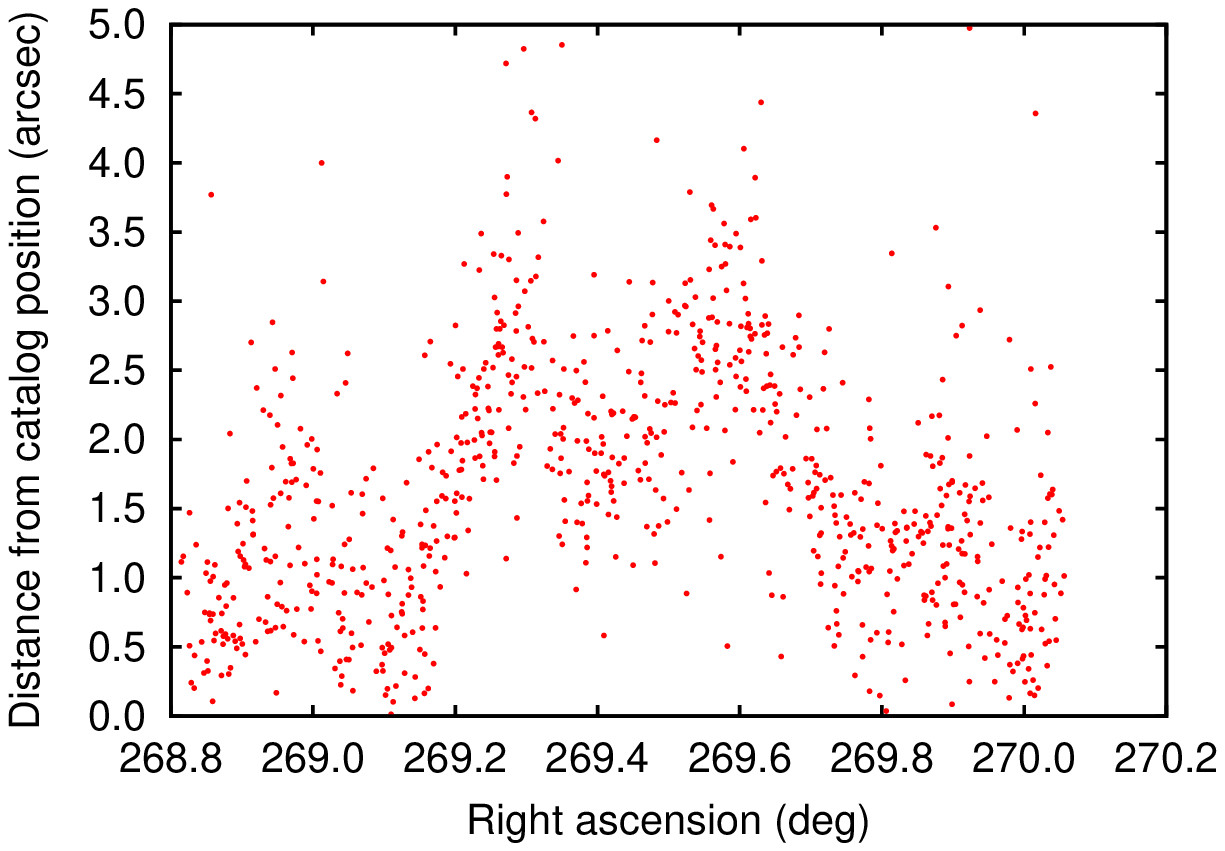}
\caption{Deviation from the catalog position as a function of R.A. Plate solution with the 2nd order
polynomial correction is applied for this $1.\!\!^\circ3 \times
1.\!\!^\circ3$ field digitized with the Expression~11000XL scanner.}
\label{fig:saw}
\endminipage\hfill
\minipage{0.48\textwidth}
\includegraphics[width=1.0\textwidth]{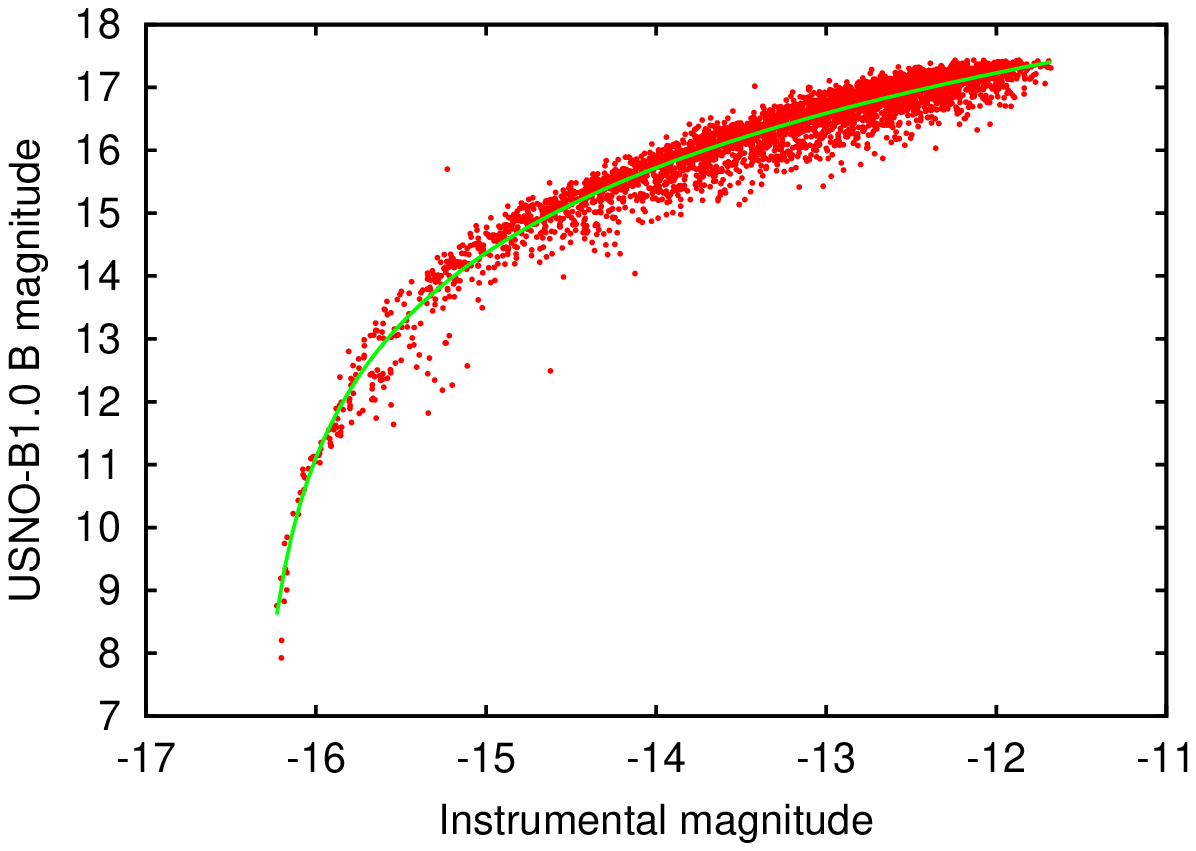}
\caption{The magnitude calibration curve for the same field as
Fig.~\ref{fig:saw}. Points represent stars matched with
the USNO-B1.0 catalog.}
\label{fig:calib}
\endminipage
\end{figure}

The TIFF images are converted into FITS format using the
\texttt{tiff2fits}\footnote{\url{ftp://scan.sai.msu.ru/pub/software/tiff2fits/}}
code. Data from only the green channel are used to write a monochrome FITS
image. The negative images are inverted at this stage to have white stars on
dark background.
Large images are cut into pieces of about $1^\circ \times 1^\circ$ so
sensitivity variation caused by vignetting and other aberrations in the
astrograph's optics as well as atmospheric transparency variations can be approximated as a
linear function of an object's position on a small image. Since plates
that belong to the same ``field'' may have offsets of more than $1^\circ$ between
their centers, a star is used as a reference point for cutting to ensure that 
the same sky area is covered by small images resulting from cutting scans of different plates.

Each series of small images is processed independently using the
\texttt{VaST}\footnote{\url{http://scan.sai.msu.ru/vast/}} variability
search software \cite{2005ysc..conf...79S}. \texttt{VaST} is using
\texttt{SExtractor}\footnote{\url{http://www.astromatic.net/software/sextractor}}
\cite{1996A&AS..117..393B} to perform object detection and aperture
photometry (the aperture size is determined individually for each image to
compensate for seeing variations) and performs cross-identification of stars
detected on the images producing lightcurves of all detected stars as an
output. The circular aperture size is optimized for measuring stars with $B>13$.
The images are plate-solved using the 
\texttt{Astrometry.net}\footnote{\url{http://astrometry.net/}} software
\cite{2008ASPC..394...27H,2010AJ....139.1782L} and the internal magnitude
scale is calibrated by matching the detected stars to the USNO-B1.0 catalog
\cite{2003AJ....125..984M}. Following \cite{2010AJ....140.1062L} we use the
relation of the form $ m_1 = a_0 \times \log_{10}\left({10^{a_1 \times (m_2 - a_2)} + 1}\right) + a_3$
proposed by \cite{2005MNRAS.362..542B} to match catalog $B$ ($m_2$) magnitudes
to the measured aperture magnitudes ($m_1$) through the fitted coefficients
$a_0,a_1,a_2,$ and $a_3$ (Fig.~\ref{fig:calib}). This relation is also
utilized to match instrumental magnitude scales of individual frames before
performing absolute calibration.
The obtained lightcurves are used to search for variable stars using an
RMS--magnitude plot and period search techniques \cite{2010ARep...54.1000K}.

\section{Results}

\begin{figure}
\centering
\includegraphics[width=0.75\textwidth]{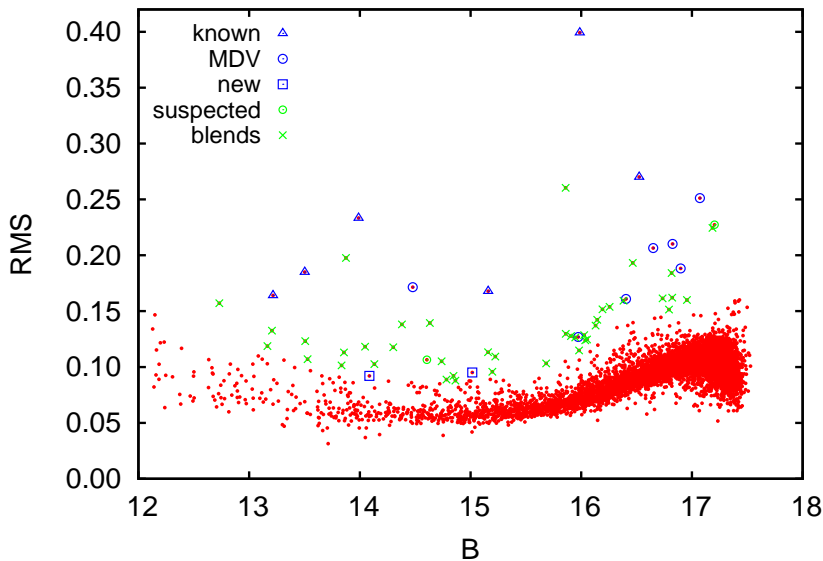}
\caption{Lightcurve RMS as a function of mean magnitude.}
\label{fig:data_msigma}
\end{figure}

\begin{figure}
\minipage{0.48\textwidth}
\includegraphics[width=1.0\textwidth]{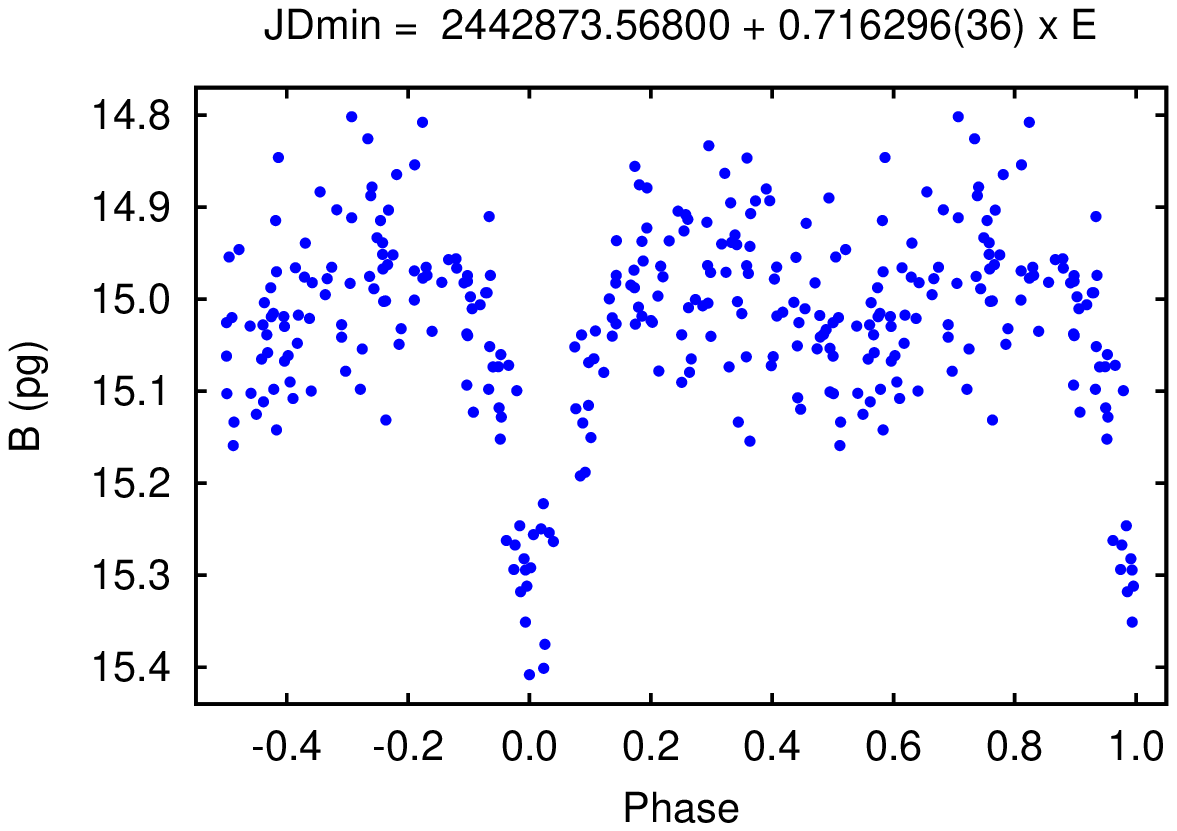}
\caption{Lightcurve of B1.0~0953-0319502.}
\label{fig:new01}
\endminipage\hfill
\minipage{0.48\textwidth}
\includegraphics[width=1.0\textwidth]{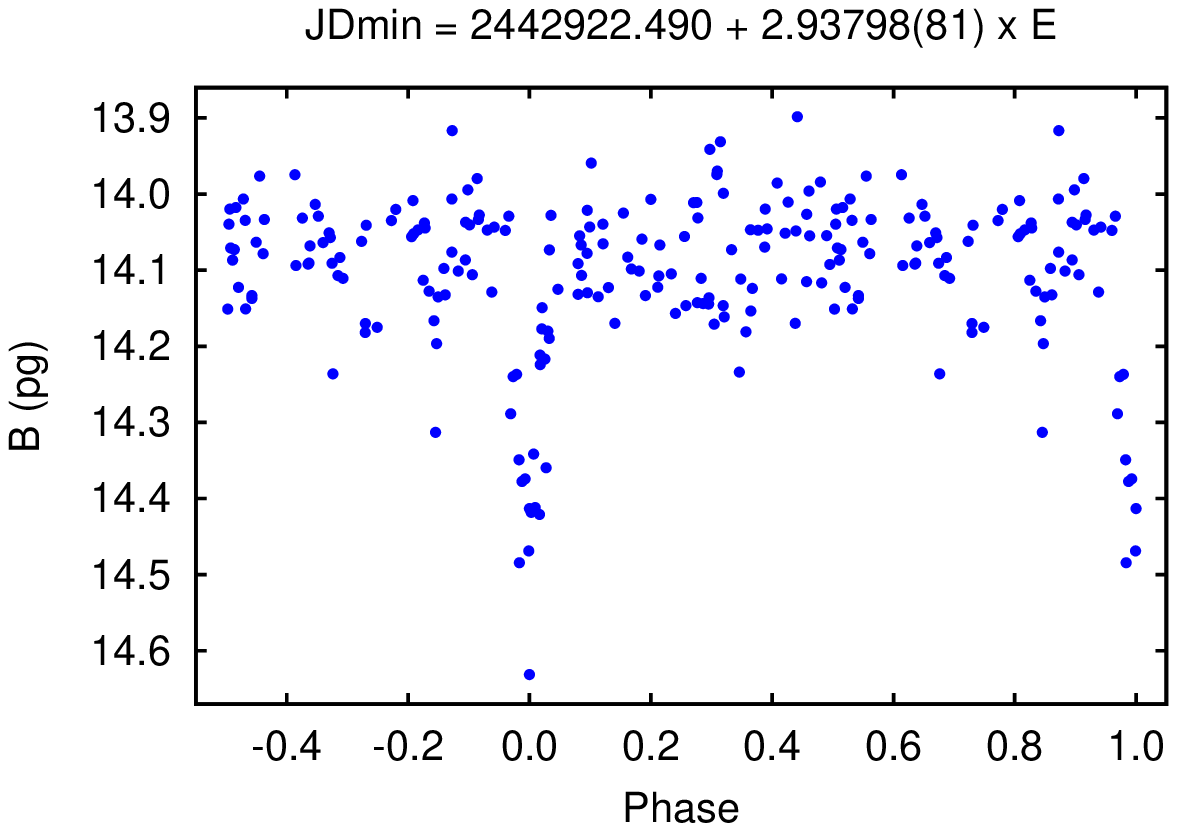}
\caption{Lightcurve of B1.0~0944-0313124.}
\label{fig:new02}
\endminipage
\end{figure}

Fig.~\ref{fig:data_msigma} presents the RMS--magnitude plot for the test
field digitized with our new Expression~11000XL scanner. The plot marks
previously known variable stars in this field (``known''), variables
identified by \cite{2008AcA....58..279K} (``MDV''), ``suspected'' variables,
and constant stars with photometry corrupted by a close neighbor
(``blends''). Two new Algol-type binaries were identified while processing
the test data (marked as ``new''), their lightcurves are presented in
Fig.~\ref{fig:new01}~and~\ref{fig:new02}.



We applied this variable stars search technique to three 
$10^\circ \times 10^\circ$ fields centered at 66\,Oph (254~plates exposed in
1976--1995), BD+60$^\circ$636 (182~plates, 1949--1989), and $\beta$\,Cas
(391~plates, 1964--1994).
Processing of 66\,Oph and BD+60$^\circ$636 fields resulted in discovery of
557 previously unknown variable stars including 6~Cepheids (Type~I and II), 
147~RR~Lyrae type variables, 12~High-amplitude $\delta$~Scuti
stars (HADS), 168~red semiregular and irregular (types SR and L)
variables, 222~eclipsing binaries, 2~BY~Dra type stars. 
Preliminary processing of 50\,\% of
the $\beta$\,Cas field (the most well-sampled of the three studied fields)
resulted in detection of 604~variable stars (454~of them new) among  
$\sim 51000$ stars in the magnitude range accessible for our variability
search. We estimate that 
$1.2 \pm 0.1$\,\% of the stars show easily-detectable (amplitude $>0.3m$)
light variations; $0.7 \pm 0.1$\,\% of the stars are eclipsing binaries 
($64 \pm 4$\,\% of them are EA type, $22 \pm 2$\,\% are EW type and 
$14 \pm 2$\,\% are EB type); $0.3 \pm 0.1$\,\% of the stars are red variable
giants and supergiants of M, SR and L types. The fraction of pulsating
variable stars of all types is expected to be a strong function of 
the Galactic latitude and deserves a more detailed investigation.
The errors are estimated from the Poisson statistics and cannot account for
any systematic effects remaining in our search.

%
%
%
%
%
%
%
%
%
%
%
%
%
%

{\small {\bf Acknowledgements.} This work is supported by the RFBR grant 13-02-00664.}

\bibliographystyle{Science}
\bibliography{sokolovsky_var_star_search}

\end{document}